\documentclass[amsmath,
  amssymb,aps,prd,nofootinbib,
reprint,
twocolumn,
superscriptaddress,altaffilsymbol]{revtex4-1}

\usepackage{amsmath,mathtools}
\usepackage{amsfonts}
\usepackage{amssymb}
\usepackage{graphicx}
\usepackage{psfrag}
\usepackage[hang]{footmisc}
\usepackage{setspace}
\newcommand*{\kth} {\ensuremath{{k_{\rm th}}}}

\newcommand{\Rinst}{{{R}_{\rm inst}}}

\newcommand{\beq}{\begin{equation}}
\newcommand{\eeq}{\end{equation}}
\newcommand{\bea}{\begin{eqnarray}}
\newcommand{\eea}{\end{eqnarray}}
\newcommand{\beqn}{\begin{equation*}}
\newcommand{\eeqn}{\end{equation*}}
\newcommand{\bean}{\begin{eqnarray*}}
\newcommand{\eean}{\end{eqnarray*}}

\newcommand*{\cref}[1]{Chapter~\ref{#1}}

\begin{document}

%\begin{flushright} {\footnotesize IC/2007/001
%\\ HUTP-07/A0002}  \end{flushright} 

\title{Collapse threshold for a cosmological Klein Gordon field}
\author{Juan Carlos Hidalgo}
\email{Corresponding author: hidalgo@fis.unam.mx}
\affiliation{Instituto de Ciencias F\'{\i}sicas, Universidad Nacional 
Aut\'onoma de M\'exico, Cuernavaca, Morelos, 62210, Mexico}
\author{Josu\'e De Santiago}
%\email{marco.bruni@port.ac.uk}
\affiliation{Instituto de Ciencias F\'{\i}sicas, Universidad Nacional
Aut\'onoma de M\'exico, Cuernavaca, Morelos, 62210, Mexico}
\author{Gabriel German}
%\email{hidalgo@fis.unam.mx}
\affiliation{Instituto de Ciencias F\'{\i}sicas, Universidad Nacional
Aut\'onoma de M\'exico, Cuernavaca, Morelos, 62210, Mexico}
\author{Nandinii Barbosa-Cendejas}
%\email{n.barbosa.cendejas@gmail.com}
\affiliation{Instituto de Ciencias F\'{\i}sicas, Universidad Nacional
Aut\'onoma de M\'exico, Cuernavaca, Morelos, 62210, Mexico}
\affiliation{Facultad de Ingenier\'ia El\'ectrica,
Universidad Michoacana de San Nicol\'as de Hidalgo,\\Morelia, Michoac\'an, M\'exico}
\author{Waldemar Ruiz-Luna}
%\email{hidalgo@fis.unam.mx}
\affiliation{Instituto de Ciencias F\'{\i}sicas, Universidad Nacional
Aut\'onoma de M\'exico, Cuernavaca, Morelos, 62210, Mexico}
\affiliation{Facultad de Ciencias, Universidad Nacional Aut\'onoma de M\'exico, Ciudad de M\'exico, 04510, Mexico}

\date{\today}
\begin{abstract}%%%%%%%%%%%%%%%%%%%%%%%%%
%%%%%%%%%%%%%%%%%%%%%%%%%%%%%%%%
Oscillating scalar fields are useful to model a variety of matter components in the universe. One or more scalar fields participate in the reheating process after inflation, while at much lower energies scalar fields are robust dark matter candidates. Pertaining structure formation in these models, it is well known that inhomogeneities of the Klein-Gordon field are unstable above the characteristic De Broglie wavelength. In this paper we show that such instability implies the existence of a threshold amplitude for the collapse of primordial fluctuations. We use this threshold to correctly predict the cut--off scale of the matter power spectrum in the scalar field dark matter model. Furthermore, for a Klein-Gordon field during reheating we show that this same threshold allows for abundant production of structure (oscillons but not necessarily black holes). Looking at the production of Primordial Black Holes (PBHs) in this scenario we note that the sphericity condition yields a much lower probability of PBH formation at the end of inflation. Remarkably, even after meeting such stringent condition, we find that PBHs may be overproduced during reheating. We finally constrain the epochs at which an oscillating Klein-Gordon field could dominate the early universe. 

\end{abstract}
\maketitle
%%%%%%%

\section{Introduction}
Scalar fields are ubiquitous in the interpretation of dominant components of the Universe at several stages of its evolution. A single minimally-coupled and potential-dominated scalar field, the inflaton, stands as the most plausible source for the primordial accelerated expansion of the universe (see e.g. \cite{Langlois:2010xc,*Wang:2013eqj} for recent reviews). At a subsequent stage, when the kinetic energy of the inflaton becomes significant, the scalar field may develop oscillations with instabilities which produce a swift transfer of energy to the standard model of particles (see \cite{Bassett:2005xm,*Allahverdi:2010xz,Amin:2014eta} for reviews). This is one of many possibilities of the reheating process.  

At later stages, and at much lower energy scales, a coherently oscillating scalar field has proved to be a good candidate for dark matter \cite{Preskill:1982cy,*Turner:1983he,*Baldeschi:1983mq,*Matos:1999et,Hu:1998kj,Marsh:2015xka,Hui:2016ltb}. The so-called Scalar Field Dark Matter (SFDM) may dominate the matter content of the Universe from the time of matter-radiation equivalence (at redshift $z\approx 3361$) up to dark energy domination ($z \simeq 0.7$). The simplest model of Scalar Field Dark Matter is a free scalar field minimally coupled to gravity and with a light mass in its potential; the Klein-Gordon field (K-G). This provides a falsifiable model with characteristic observables. In particular, in the process of structure formation the matter perturbations show instabilities only above a characteristic scale. %, $\Rinst$. 
Such instability was first interpreted as a Jeans' instability in \cite{Khlopov:1985jw, Bianchi:1990mha} and is associated, as we shall see, to the de Broglie wavelength of the scalar field. The instability scale for inhomogeneities implies the existence of a cut-off in the matter power spectrum, which has historically provided a plausible solution to the missing satellite problem on galactic scales \cite{Kamionkowski:1999vp,BoylanKolchin:2011dk}. 

In the non-perturbative regime the growth of instabilities in oscillating field cosmologies has been mostly studied numerically (see \cite{Widrow:1993qq,*Davies:1996kp,*Coles:2002sj,*Short:2006md,Schive:2014hza,Torres:2014bpa,*Rekier:2015isa}  for the SFDM case and  \cite{Amin:2011hj,Torres-Lomas:2014bua} for the reheating scenario), confirming the perturbative result that the inhomogeneities dissipate at small scales compared with the instability scale, while at large scales the non-linear evolution of an oscillating scalar field inhomogeneitites follows closely that of cold dark matter structures. Furthermore, in the cosmological environment, the formation of virialised structures from an oscillating scalar field is expected at the perturbative level \cite{Magana:2012xe}, as well as the formation of solitonic non-linear structures (oscillons for the reheating fields, or boson stars in the case of SFDM). 

In this paper we show that the presence of an instability scale for perturbations of an oscillating scalar field, $\Rinst$, implies the existence of a non-vanishing threshold amplitude for the collapse of primordial perturbations. We find such threshold with the aid of the spherical collapse model, where the maximum radius of an overdensity is related to the amplitude of the density contrast. Applying the three-regions model of spherical collapse \cite{Kopp:2010sh,Harada:2013epa} to the universe dominated by a K-G field we derive analytic expressions for the threshold amplitude required for gravitational collapse. 

{For the case of SFDM, the De Broglie wavenumber determines the cut--off scale in the linear regime of the matter power spectrum. This prediction normalises the spherical collapse criterion to determine the amplitude threshold of inhomogeneities and we extend the cut--off value into the non-linear scales.}

%% In the context of the oscillating inflaton at the reheating stages we compute the abundance of primordial black holes and use the observational bounds (including or not the sphericity condition) to constrain the energy scales at which this field may have dominated the universe.
%%
%%We exemplify the validity of this threshold amplitude by correctly approximating the cut-off scale of the power spectrum in a SFDM scenario.
In the reheating scenario, we compute the {non-linear} threshold to form structure after the end of inflation and find that most {primordial} inhomogeneities should collapse even if we extrapolate the observed (red) spectrum to the smallest scales. This is interpreted as an efficient production of oscillons cite{Okawa:2013jba}. However, looking at the formation of primordial black holes (PBHs) in the reheating era, our result leads us to adopt a more conservative approach by considering the sphericity condition of primordial fluctuations to collapse into a black hole \cite{Khlopov:1980mg,*Khlopov:1982ef,Harada:2016mhb}. We find that despite such restrictions PBHs could be overproduced in light of the known bounds to the PBH abundance (e.g. \cite{Carr:2009jm}).
%\cite{Clark-1612.07738}. 
In turn, we argue that the production of PBHs limit the energy scales at which an oscillating reheating scalar field permeates the early universe {and compute bounds to thermalisation values for reheating models via an oscillating scalar field}.

The paper is organised as follows: In section \ref{sec:II} we review the spherical collapse model and discuss Carr's criterion for the existence of a threshold amplitude of collapse in barotropic fluids  \citep{Carr:1975qj}. In section \ref{sec:III} we identify the unstable regime of an oscillating Klein-Gordon field and adapt Carr's criterion to determine the threshold amplitude required for scalar field inhomogeneities to collapse at linear and non-linear level. In section \ref{sec:IV} we look at scenarios where the derived threshold finds an application as described above. We discuss our results and sketch directions of future work in the final section \ref{sec:V}. 

%%Taking on account the instability scale arising at the perturbative regime, we derive the threshold amplitude of primoridal density perturbations required for the collapse which, in the case of an axion field, yields a well known feature in the matter power spectrum. After checking for consistency with the power spectrum cut-off, we turn to the inflaton field at the reheating stages to compute the abundance of objects collapsing at each energy scale in the early universe. Interpreted as particles of dark matter in the form of ultra compact mini haloes (UCMH) or primrodial black holes (PBHs), we show finally how the constraints on abundance of collapsed structures yield an observational bound on the scales allowed for the simplest model of reheating. 
%%%%%%%%%%%%%%%%%%%%%%%%%%%%%%%%
%%%%%%%%%%%%%%%%%%%%%%%%%%%%%%%%
\section{Spherical Collapse}
\label{sec:II} 
The spherical collapse model is a useful tool to characterise the fate of inhomogeneities in cosmology (see \cite{Gunn:1972sv,Lokas:2000cn} for seminal works). In this model, a spherically symmetric overdensity corresponds to a positive curvature region, embedded in a flat background. The non-linear evolution of these inhomogeneities provides information of the collapse (or virialisation) times for dark matter/dark energy models. This is useful to discriminate models via the abundance of
collapsed (virialised) objects as a function of redshift (e.g.~\cite{Lahav:1991wc,Pace:2010sn,LoVerde:2014rxa}). On the
other hand, for suitable matter contents one can compute the range of amplitudes of primordial fluctuations that undergo a complete gravitational collapse and get to form black holes (\cite{Musco:2004ak,*Musco:2012au,*Bloomfield:2015ila})
%(e.g. \cite{Carr:1975qj,Kopp:2010sh,Harada:2013epa}).  

The simplest version of the spherical collapse model is the Top-Hat
model, and from it we can derive basic equations, useful for our analysis\footnote{Most of this section is a synthesis of the thorough analysis developed in \cite{Kopp:2010sh,Harada:2013epa}.}. This model considers the homogeneous and isotropic Friedmann-Lemaitre-Robertson-Walker (FLRW) metric,  
\beq
\label{FLRW:metric}
ds^{2} = -dt^{2} + a^{2}(t)\left(\frac{dr^{2}}{1-Kr^2} +r^{2}d\Omega\right),
\eeq

\noindent where $r$ is the comoving areal radius and $d \Omega$ accounts for the differential angular displacement. $K$ is the curvature characterising a closed ($K=1$), flat ($K=0$), or open ($K=-1$) universe. An alternative expression for the metric of the closed universe is given in terms of the comoving radial coordinate $\chi$ as 
\beq
ds^2 = -dt^2 +a^2(t)\left( d\chi^2 + \sin^2\chi d\Omega\right).
\eeq

\noindent %with the areal radius given by $R_a = a(t) \sin(\chi)$.
The radial coordinate $\chi$ can adopt values in the range $[0,\pi]$ and we denote the maximum comoving radius of the closed universe as $\chi_a$. The evolution of the scale factor $a(t)$ is given by the Friedmann equation, which in terms of the Hubble factor, $H \equiv d ( \ln a) / d t$, is written as
\beq
\label{friedmann:eq}
H^2 = \frac{\kappa}{3}{\rho} - K\frac{c^2}{a^2},
\eeq

\noindent where ${\rho}$ accounts for the homogeneous matter
density of the universe, and where $\kappa = 8\pi G$. 

A Top Hat universe configuration, constituted by an overdense central region represented by a closed universe (described by quantities labelled with subindex $a$) and surrounded by a flat background universe (whose quantities are denoted with subindex $b$). By definition the matter densities are related through the overdensity $\delta \rho$, that is,
\beq
\label{delta:def}
\rho_a = \rho_b + \delta \rho = \rho_b (1 + \delta),
\eeq

\noindent which implicitly defines the density contrast $\delta$. As an initial condition, at time $t_i$, we demand both regions to expand at the same rate, then $H_a(t_i) = H_b(t_i)$ and   
\bea
H_b^2 &=& \frac{\kappa}{3}{\rho_b} =  \frac{\kappa }{3}{\rho_b(1 +
  \delta)} -  \frac{\kappa}{3}{\delta \rho} \notag\\
  &=&  \frac{\kappa
}{3}{\rho_a} - K_a\frac{c^2}{a(t_i)^2} = H_a^2.   
\label{Ha:Hb}
\eea

Identifying terms in the last equality with those describing a closed universe in Eq.~\eqref{friedmann:eq}, the positive curvature finds a natural interpretation in terms of the matter overdensity. A more suitable interpretation of the above is to adopt the conventional unit value for the curvature $K_a = 1$ and demand a uniform expansion at all times within the perturbative regime. This condition defines a gauge for the matter perturbation, the uniform expansion or uniform Hubble gauge, and implies that
\beq
\frac{8 \pi G}{3} \delta \rho_{UH} = H_a^2
\delta_{UH} = \left(\frac{c}{a(t)}\right)^2 ,
\eeq

\noindent where the subscript ${}_{UH}$ denotes the uniform expansion gauge. We can thus express the matter density contrast in terms of the areal radius of the overdense region $R_a = a(t) \sin(\chi_a)$ and the Hubble radius $R_{H} = c/H$, as
\beq
\label{delta:UH}
\delta_{UH} = \left(\frac{R_H}{R_a}\right)^2 \sin^2 \chi_a.
\eeq

\noindent Finally, evaluating the above at the time of horizon crossing one finds
\beq
\label{delta:chi1}
\delta_{\rm UH}^H = \sin^2 \chi_a.
\eeq

One of the limitations of the Top-Hat model is that the total density is increased by the addition of the fluctuation $\delta \rho$ without compensation, in a universe which total average density should match the background $\rho_b$. This issue is addressed in the three-regions model here described, and developed by Kopp, \textit{et al.}~\cite{Kopp:2010sh}. This provides a suitable framework for the collapse of primordial fluctuations with null contribution, on average, to the background density. The internal patch represents the overdense region, with density $\rho_+$ and positive spatial curvature, surrounded by an under-density that compensates the matter contribution to match the average density ${\rho_b}$ of a third, background flat universe.
%In the
%homogeneous expansion, or uniform Hubble gauge, the matter density
%contrast $\delta^{UH}= (\rho_+ - \bar{\rho}) / \bar{\rho}$ of the
%overdense region is still related to the comoving radius of the overdense
%region $\chi_{+}$ as in \eqref{delta:chi1}, 
%\beq
%\label{delta:chi}
%\delta^{UH}_H = \sin^2 (\chi_{+}).
%\eeq\

The evolution of the overdense region is equivalent to the solution
to the scale factor in a closed Friedmann universe, with the maximum expansion reached beyond the perturbative regime when 
$H_a(t_{\rm max}) = 0$. At maximum expansion, the scale factor can be written as
\bea
a_{\rm max} &=&\sqrt\frac{3c^2}{\kappa \rho_{a}(t_{\rm max})} =
\left(1 - \frac1\Omega_a\right)^{-1}a(t) \notag\\ &=& \frac{\Omega_a}{(\Omega_a -
  1)^{3/2}}\frac{c}{H},   \label{a:max}
\eea

\noindent where $\Omega_a =\kappa\rho_a(t)/3 H^2_a$ is the matter density parameter in the overdense region evaluated
at a time $t < t_{\rm max}$. At maximum expansion,  the
maximum areal radius of the overdensity is related to the  
comoving radius through the scale factor
%in units of the Hubble radius 
\bea
%\label{Rmax:amax}
R_{\rm max} &=& a_{\rm max} \sin(\chi_a), \notag\\
\label{Rmax:chi}
R_{\rm max} &=& \frac{\Omega_a}{(\Omega_a - 1)} R_a(t).
\eea

Finally, using Eq.~\eqref{delta:chi1} and writing the particle horizon radius $R_{H_a} = c/H_a$ we arrive at the relation between the matter density perturbation and the radius of maximum expansion,
\beq
\label{deltaUH:Rmax}
\delta_{\rm UH}^H = \left(\frac{R_{\rm max}}{a_{\rm max}}\right)^2 = (\Omega_a
-1) \frac{R_a^2}{R_{H_a}^2}
\eeq

{
Let us now use this result to identify cosmologically unstable configurations. Carr's argument for gravitational collapse \cite{Carr:1975qj} can be interpreted in the three-regions collapse as follows \cite{Harada:2013epa}: 
If the radius of maximum expansion $R_{\rm max}$ lies above a characteristic instability scale $\hat{ R_{\rm J}}$ (a suitable rescaling of the Jeans' length in a perfect fluid), the inhomogeneity will be gravitationally unstable and collapse. Mathematically,
\beq
\label{Rmax:f}
 R_{\rm max}/ a_{\rm max}  > \hat{R_{\rm J}}/ a_{\rm max} = f,
\eeq
}

{where the rescaling factor is chosen such that \cite{Carr:1975qj,Harada:2013epa}
\beq
\hat{R}_{\rm J} = \frac{3}{8 \pi^2} R_{\rm J}(t_{\rm max})= \sqrt{w}a_{\rm max},
\eeq
}
{\noindent and therefore
\beq 
f = \frac{1}{a_{\rm max}}\frac{c_s}{\sqrt{8\pi G\rho_{\rm max}}} = \sqrt{w},
\eeq
}
Eq.~\eqref{deltaUH:Rmax} indicates that the condition for the instability scale in Eq.~\eqref{Rmax:f} implies the existence a threshold value for the density contrast at the time of horizon crossing in the uniform expansion gauge, that is
\beq
\label{crit:delta}
f^2 = w < \delta^{UH}_H < 1. 
\eeq

\noindent Here the upper limit is imposed by the maximum ratio of the areal to comoving radius at $\chi_a = \pi/2$ (We take values in the range $0 <\chi_a < \pi/2$ that represents configurations of type I according to the classification of \cite{Kopp:2010sh}; the value of the areal radius at $\chi_a = \pi$ is zero and this denotes a separate universe configuration).
% It is thus straightforward to constrain the amplitude of a collapsing fluctuation at horizon crossing
%\beq
%\label{crit:zeta}
%-2\ln \left[ \cos \left( \frac{\arcsin \sqrt{f}}{2}\right) \right] < \zeta \leq \ln 2.
%\eeq

In \citep{Harada:2013epa} a refinement of this criterion is proposed for the case of barotropic fluids with equation of state $w$. It consists on a precise determination of the non-linear instability scale derived from the balance of the propagation time of a sound wave in the radius of an initially expanding overdensity, and the collapse time of the overdensity. The resulting instability scale at the maximum expansion is 
\beq
R_{\rm inst} = a_{\rm max}\sin\left(\frac{\pi\sqrt{w}}{1 + 3 w}\right).	
\eeq

\noindent {This new value corrects the linear instability scale (Jeans length) $\hat{R_{J}} = a_{\rm max}\sqrt{w}$}. The corresponding threshold amplitude for collapse changes from $\delta_c = {w}$ to the more precise form
\beq
\delta_c = \left(\frac{R_{\rm inst}}{a_{\rm max}}\right)^2 = \sin^2 \left(\frac{\pi\sqrt{w}}{1 + 3 w}\right).
\eeq

\noindent This accurately reproduces the numerical value derived from
simulations of Primordial Black Hole formation
\cite{Musco:2004ak,*Musco:2012au,*Bloomfield:2015ila}. Note that the
improved prescription for the instability scale can be expressed in
terms of the ratio $ f $. We can
generalize the prescription for the threshold amplitude in Eq.~\eqref{crit:delta} to the form:
\beq
\label{R:inst}
\delta_c= \left( \frac{R_{\rm inst}}{a_{\rm max}} \right)^2 = \sin^2 \left(\frac{\pi f}{1 + 3 f^2}\right). 
\eeq

\noindent We interpret this last equation as the correspondence between the perturbative and the non-perturbative instability scales in the three-regions spherical collapse model. The critical value for the collapse fixes a critical value for the coordinate radius in virtue of
Eq.~\eqref{delta:chi1}  
\beq
\label{chi:f}
\chi_c = \frac{\pi  f}{1 + 3 f^2}.
\eeq

\noindent In the following we apply this prescription
to derive the instability threshold amplitude of scalar field inhomogeneities. 

For completeness, let us end this section presenting the relation
between the comoving radial coordinate and the average
value of $\zeta$, the curvature perturbation in the comoving density
gauge \cite{Kopp:2010sh}. The threshold amplitude $\bar{\zeta}_c$ is related to the comoving radius as, 
\beq
\label{barzeta:chi}
\bar\zeta_c = \frac13 \ln\left[ \frac{3 \chi_c - \sin(\chi_c)\cos(\chi_c)}{2
    \sin^3(\chi_c)}\right].
\eeq

\noindent We shall use this value to test the collapse of primordial curvature
inhomogeneities.

%%%%%%%%%%%%%%%%%%%%%%%%%%%%%%%%
%%%%%%%%%%%%%%%%%%%%%%%%%%%%%%%%
\section{Amplitude for collapse of an oscillating scalar field}
\label{sec:III}
While the spherical collapse model is exact only for the case of a pressureless fluid, 
the three-regions model of spherical
collapse has been successfully employed to derive the threshold
amplitudes for barotropic fluids with non-vanishing equations of
state \cite{Harada:2013epa}. The case of an oscillating scalar field
must be treated separately.  A naive fluid interpretation of the scalar field would lead to the erroneous
conclusion that Jeans length exists and is either zero (if determined from the equation of state) or equal to the Hubble radius (if guided by the adiabatic sound speed), and either all or no scales 
and amplitudes should collapse (the incompleteness of a fluid description is discussed in \cite{Christopherson:2012kw}). 
The gravitational instability of perturbations around the cosmological solution has a different nature as we shall see here.

The evolution equation for the Klein-Gordon field is that of a minimally-coupled canonical scalar field with potential $V=  m^2 \varphi^2/2$:
 \begin{equation}
 \label{KG:eqn}
g^{\alpha \beta}\varphi_{;\alpha \beta} -m^2 \varphi = 0.
\end{equation}
In a flat FLRW space-time, the homogeneous field thus obeys the equation
\beq
\label{KG-field}
\ddot{\varphi} + 3H\dot{\varphi} + m^2 \varphi= 0.
\eeq  

\noindent Our interest is in the regime 
in which the dynamical time for the field is much smaller than the cosmological
expansion rate ($ m \gg H$). As a result, the solution to Eq.~\eqref{KG-field}
is given by 
\beq
\label{bg:phi}
\varphi (t) = \varphi_0 a^{-3/2}\left[
  \sin(mt) + \mathcal{O}\left\{\left(\frac{H}{m}\right)^2 \right\}\right].
\eeq

\noindent In the homogeneous space-time, the energy-momentum tensor in the
comoving frame can be matched to that of an homogeneous perfect
fluid. In the proper frame, the components of the scalar field
stress-energy tensor are interpreted as the homogeneous density
$ \rho_{\varphi} = - T^0_0 =  \dot{\varphi}^2/{2} + m^2
\varphi^2 /{2}$ and isotropic pressure
$ p_{\varphi}=   T^j_j/3 =  \dot{\varphi}^2/{2} - m^2
\varphi^2/{2} $.
When we average the above solution over a single period of oscillation
$1/m$, we find that
\bea
\langle \rho\rangle & = & \frac{1}{2}\langle \dot{\varphi}^2\rangle
+\frac{1}{2}m^2 \langle \varphi^2\rangle \approx m^2 \langle
\varphi^2\rangle,
\notag \\
\Rightarrow &&\langle \rho\rangle = \frac{\varphi_0^2}{2} m^2 a^{-3}
+ \mathcal{O}\left\{\left(\frac{H}{m}\right)^2\right\}.
\label{avg:rho}
\\
\langle p \rangle & = & \frac{1}{2}\langle \dot{\varphi}^2\rangle -
\frac{1}{2}m^2 \langle \varphi^2 \rangle \approx
\frac{9\varphi_0^2H^2}{16\,a^3}, \notag \\ 
\Rightarrow &&\langle p \rangle = \mathcal{O}\left\{a^{-6}\right\}.
\label{avg:p}
\eea

\noindent As a consequence of this behavior, if the oscillating
scalar field dominates the Universe for sufficiently long time, it effectively
behaves as pressureless dust in the background. 

In the perturbative regime, field fluctuations present an instability
scale explicit in the evolution equation of the (modified)
Mukhanov-Sasaki variable $\nu$ \cite{Mukhanov:1988jd,Sasaki:1986hm},
but not so evident in the evolution equation of the perturbed K-G field \eqref{KG:eqn}. 
This function can be written in terms of the curvature perturbation $\zeta$ as 
% relacionada a $\zeta$ por $-\sqrt{2\kappa}\nu=\mu_s =-2a
%\sqrt{\gamma}\zeta$, donde
% $\gamma = 1-\frac{\mathcal{H}'}{\mathcal{H}^2}$.
\beq
\hat{\mu} \equiv - 2 \sqrt{a^3 \left(
1-\frac{\mathcal{H}'}{\mathcal{H}^2}\right)}\zeta\,,
\eeq 

\noindent where $\mathcal{H}$ is the Hubble factor defined in conformal time. 
%\footnote{Note that the linear 
%relation between the comoving curvature perturbation and the 
%Mukhanov-Sasaki variable ensures that the instability wavenumber 
%is common to both quantities at the perturbative level.} 
The evolution equation for this variable is, in Fourier space \cite{Mukhanov:1990me}, 
\bea
&&\ddot{\tilde{\mu}}_s+\Bigg\{\frac{k^2c^2}{a^2}
+\frac{d^2  V}{d\varphi^2}+3\kappa \dot{\varphi}^2
-\frac{\kappa^2}{2H^2}\dot{\varphi}^4\notag \\
&+&\frac{3\kappa}{4}\left(\frac{\dot{\varphi}^2}{2}-V\right)
+2\kappa \frac{\dot{\varphi}}{H}\frac{dV}{d\varphi}\Bigg\}\tilde{\mu}_s =0.
\label{MS:eqn}
\eea

\noindent Inserting the solution \eqref{bg:phi} for the background
field and its quadratic potential, the reduced equation, up to order $\mathcal{O}(a^{-3})$, is 
\begin{align}
&\big[\frac{k^2c^2}{a^2} + m^2 +  \\ & 2\sqrt{6\kappa}m^2 \varphi_0 a^{-3/2} \sin(mt) \cos(mt)\big]\tilde{\mu}_s
+\, \ddot{\tilde{\mu}}_s=0.
\notag
%\label{MS:redeqn}
\end{align}

\noindent Finally, with the aid of trigonometric identities and the change of variable $z= m t + \pi/4$, we arrive at an expression
close to a Mathieu equation 
\begin{equation}
\frac{d^2 \tilde{\mu}_k}{dz^2} +\left[ A_k -2q\cos (2z)\right] \tilde{\mu}_k =0,
\label{mathieu:eqn}
\end{equation}

\noindent where
\begin{equation}
q=\frac{1}{2}\sqrt{6\kappa} \varphi_0 a^{-3/2}= 3\frac{H}{m} + \mathcal{O}\left\{\left(\frac{H}{m}\right)^2\right\},
\label{cu}
\end{equation} 
and
\begin{equation}
A_k =1+\frac{k^2c^2}{m^2 a^2}.
\label{aka}
\end{equation}
The Mathieu equation has been analyzed for the oscillating K-G field in previous 
works (see e.g. \cite{Kofman:1997yn,Jedamzik:2010dq}) noting that the above values 
yield a single instability band for the perturbations inside the Hubble horizon, 
namely
\begin{equation}
1-q <A_k <1+q,\,
\leftrightarrow \,-q < \frac{k^2c^2}{m^2 a^2}< q.
\end{equation}

\noindent Together with Eq.~\eqref{cu}, this yields the instability wavenumber 
\beq
\label{k:inst}
k_{\rm lin} = \frac{a \sqrt{3mH}}{c}.
\eeq

\noindent The corresponding instability radius, which is proportional to
the de Broglie radius for the field, is given by  
\beq
\hat{R}_{\rm dB}=\frac{\eta a}{k_{\rm lin}} = \frac{\eta c}{\sqrt{3mH}},
\eeq

\noindent (were a hat denotes a rescaling of the de Broglie wavelength required for our non-linear analysis). In the perturbative regime, all scales above this radius and below $R_{\rm H} = c/ H$ are gravitationally unstable.
{Following Harada \textit{et~al.}~\cite{Harada:2013epa}, we have introduced
a constant $\eta$ to normalise the instability scale to the linear prescription of Eq.~\eqref{k:inst}. In their case 
this constant modifies the linear Jeans scale in order to recover
Carr's threshold criteria with a value given by $\sqrt{3/8\pi}$.
In our case, this constant modifies the linear de Broglie scale
and its numerical value will be determined in the next section in order to recover the correct cut-off in the linear regime of the matter power spectrum.}
In the spherical collapse model the corresponding nonlinear instability
scale from Eq. \eqref{R:inst}, at the time of maximum expansion, is given by 
\beq
\label{Rinst:dB}
R_{\rm inst} = a_{\rm max} \sin\left(\frac{\pi f_{\rm dB}}{1+ 3 f_{\rm dB}^2}\right),
\eeq

\noindent with 
\beq
\label{fdB:def}
f_{\rm dB} = \frac{R_{\rm dB}}{a_{\rm max}} =  \eta \sqrt{\frac{H_{\rm max}}{3m}}.
\eeq

\noindent where we have used Eq.~\eqref{a:max}, and where $H_{\rm max} = H_b(t_{\rm max})$. 

We here propose that, in analogy to the fluid case and in the spirit of Carr's criterion, the instability scale in Eq.~\eqref{Rinst:dB} implies the existence of a critical amplitude for the instability of scalar field inhomogeneities. 
Following the prescription of the spherical collapse model, the critical amplitude
of the average $\bar{\zeta}_c$ for the gravitational collapse of an oscillating K-G field, at horizon crossing time, is derived from
Eqs.~\eqref{Rmax:chi} and \eqref{chi:f}-\eqref{barzeta:chi}  
\beq
\label{crit:zeta:sf}
\bar\zeta_c = \frac13 \ln\left[ \frac{3 \chi_{\rm dB} - \sin(\chi_{\rm dB})\cos(\chi_{\rm dB})}{2
    \sin^3 \chi_{\rm dB}}\right], 
\eeq

\noindent where $\chi_{\rm dB} = {\pi f_{\rm dB}}/(1+ 3 f_{\rm dB}^2)$ is read directly from Eqs.~\eqref{Rinst:dB}-\eqref{fdB:def}. 

It is important to mention that the existence of an instability of the scalar field fluctuations does not imply their complete collapse and the ultimate formation of a black hole. As seen in Eq.~\eqref{avg:p}, the cosmological oscillating scalar field does not present a significant isotropic pressure, however, the formation of an apparent horizon for matter with small equation of state is conditioned by the sphericity of the primordial fluctuations. This issue was first first addressed by \cite{Khlopov:1980mg,*Khlopov:1982ef} and explored in further depth by \cite{Harada:2016mhb}. The sphericity condition and its relevance in the determination of Primordial Black Hole formation during reheating will be discussed below.
\section{Collapsed objects and formation of structure}
\label{sec:IV}
%%%%%%%%%%%%%%%%%%%%%%%%%%%%%%%%
Let us look at environments where the result of the previous section finds an application. Our results are valid in the limit $m\gg H$. There are at least two well known scenarios in which a K-G field in this limit may dominate in the history of our Universe: 

\textbf{Reheating}: the inflaton field may present a long-lasting period of coherent oscillations, while transferring its energy to the fields of the standard model. The oscillating scalar field may survive down to energy scales of order $E = 10^{7}\, \mathrm{GeV}$ \cite{Kawasaki:2008qe} (the energy scale required for thermalisation if reheating is to avoid the overproduction of gravitinos) or even at all values above the electroweak scale at $E = 100 \,\mathrm{GeV}$  if Supersymmetry is ignored \cite{Dai:2014jja}. This lies well below the typical values for the inflaton mass $m_{\rm inf} \approx 10^{16}\,\mathrm{GeV}$ and thus the condition $m_{\rm inf}\gg H$ is met. 
 
\textbf{The Scalar Field Dark Matter}\footnote{Several names have been associated to this single model: Fuzzy Dark Matter \cite{Hu:2000ke}, Wave Dark Matter($\Psi$DM) \cite{Bray:2012qb,Schive:2014hza}, Bose-Einstein Condensate Dark Matter \cite{Suarez:2013iw,Matos:2008ag} and Axion- or Ultra-light Axion Dark Matter (see a review in \cite{Marsh:2015xka}).}: A single Klein-Gordon field with mass in the range $m_{\rm DM} = 10^{-24}- 10^{-22}\,\mathrm{eV}$ could play the role of dark matter. Dark matter dominates the Universe from the time of matter-radiation equivalence, at $z_{\rm eq} \approx 3361$, when $H_{\rm eq} \approx 10^{-29} eV$. Thus, in the relevant epochs $m \gg H$ and the conditions leading to the threshold of collapse are met. 

We shall first look at the consequences of our collapse threshold for the formation of large scale structure in the scalar field dark matter scenario, and subsequently look at the probability of Primordial Black Hole (PBH) formation in the reheating scenario.

%%%%%%%%%%%%%%%%%%%%%%%%%%%%%%%%
\subsection{The cut-off in the power spectrum of Scalar Field Dark Matter}
\label{sec:ivA}
%%%%%%%%%%%%%%%%%%%%%%%%%%%%%%%%

For more than three decades, the Klein-Gordon scalar field (K-G field)has been a standing candidate for dark matter (Scalar Field Dark Matter-SFDM). The gravitational stability of the K-G field has been addressed as a problem for static configurations \cite{Khlopov:1985jw} or from the evolution of linear cosmological perturbations \cite{Hu:1998kj,Matos:2000ss} and only few studies have considered the non-linear collapse of the K-G field \cite{Magana:2012xe}. More recently, detailed numerical studies at the perturbative level \cite{Urena-Lopez:2015gur}, as well as through numerical simulations of the full GR spherically-symmetric system \cite{Torres:2014bpa,*Rekier:2015isa,Alcubierre:2015ipa} and N-body cosmological simulations \cite{Schive:2014hza} have been performed.

A characteristic feature of this model is a cut-off in the spectrum of density perturbations, which stems from a non-zero instability scale of the cosmological K-G field. This feature helps to alleviate the ``missing satellite'' problem by down turning the matter power spectrum at small scales \cite{Kamionkowski:1999vp, BoylanKolchin:2011de}. Let us show how the threshold value for gravitational instability derived above predicts, to a good approximation, the SFDM cut-off not previously derived analytically. 

The average amplitude for primordial curvature fluctuations in Fourier space is parametrized by a power law, with the amplitude $A_s$ and the spectral index $n_s$, as 
\beq
\label{primordial:zeta}
\zeta^2_{\rm prim}(k) =\mathcal{P}(k) =A_s \left( \frac{k}{k_0}\right)^{n_s -1}.
\eeq

\noindent {Where $k_0=0.05\rm{Mpc}^{-1}$, and} observations from the Planck satellite yield $A_s ^{1/2} = 4.63 \times 10^{-5}$  and  $n_s = 0.9681$, as the best fit values for these parameters \cite{Ade:2015xua}. When we equate this amplitude to the threshold value in \eqref{crit:zeta:sf} for a given mass of the K-G field, we obtain the threshold wavenumber $k_{\rm th}$, corresponding to the scale of the smallest fluctuations allowed to collapse, as well as the cosmic time ($1/H= a/k_{\rm th}$) when they enter the Hubble horizon (in the radiation era) \footnote{When adapting our analytic prescription for the threshold $\bar\zeta_c$, we must take on account the dominating background component and consider the factor $\Omega_m$ in equations like \eqref{cu} evaluated at horizon crossing.}. An analytic approximation to the value of the threshold wavenumber (for small $\zeta$) is given by:
\begin{equation}
\label{k:th1}
\kth = \left[ \frac{900 \eta^4}{\pi^4}\frac{\rho_{r0}}{3M_{Pl}^2} m^2 A_s  k_0^{1-n_s} \right]^{1/(5-n_s)} \,.
\end{equation}

\noindent This is simplified if we introduce the central values from Planck 2015 \cite{Ade:2015xua} to
\begin{equation}
\label{k:th2}
  \frac{\kth}{0.0687\,\rm{Mpc}^{-1}} = \left( \frac{m}{10^{-22} \rm{eV}}  \right)^{2/(5-n_s)} \eta^{4/(n_s-5)} \,.
\end{equation}

\noindent {Now we proceed to determine $\eta$ in order to have a closed analytic expression for the threshold scale.
To this purpose we use the linear instability scale (\ref{k:inst}) which at horizon crossing should be
equal to $k_h=aH$. This gives an expression for this scale as a function of the scalar field mass 
if we assume that the horizon crossing occurs within the radiation dominated era},
\begin{equation}
k_{\rm lin} = \sqrt{3m}\left(\frac{8\pi G}{3} \rho_{r0}\right)^{1/4}\,.
\end{equation}
Substituting the value of the radiation density at the present epoch we find
\begin{equation}\label{cortelineal}
\frac{k_{\rm lin}}{1.006 \rm{Mpc}^{-1}} = \sqrt{\frac{m}{10^{-22}\rm eV}} \,.
\end{equation}
\noindent This can be used to obtain the value of $\eta$ if we demand the non-linear scale to be equal to the
linear scale in the limit when $m=10^{-26}$eV and $k_{\rm lin} \approx 0.1 \rm{Mpc}^{-1}$.
We obtain $\eta=6.8\times 10^{-3}$ and the expression (\ref{k:th2}) simplifies to
\begin{equation}
\label{k:thl}
\frac{\kth}{9.7\,\rm{Mpc}^{-1}} = \left( \frac{m}{10^{-22} \rm{eV}}  \right)^{2/(5-n_s)}
\end{equation}

\noindent From this we obtain a series of threshold values for the range of masses presented in Table I. This set of values are close to those resulting from Hu \textit{et.~al.}~\cite{Hu:2000ke} in the range of validity of the model (the latter itself consistent with numerical simulations of this feature \cite{Urena-Lopez:2015gur}), for which the power drops to zero with respect to the matter power spectrum in
the $\Lambda$CDM model at the threshold values given by

\begin{equation}
\label{k:th3}
  \frac{k_{\rm Hu}}{6.5 \,\rm{Mpc}^{-1} } = \left( \frac{m}{10^{-22} \rm{eV}}  \right)^{4/9} \,.
\end{equation}

\begin{figure}[tb]
\includegraphics[width=8cm]{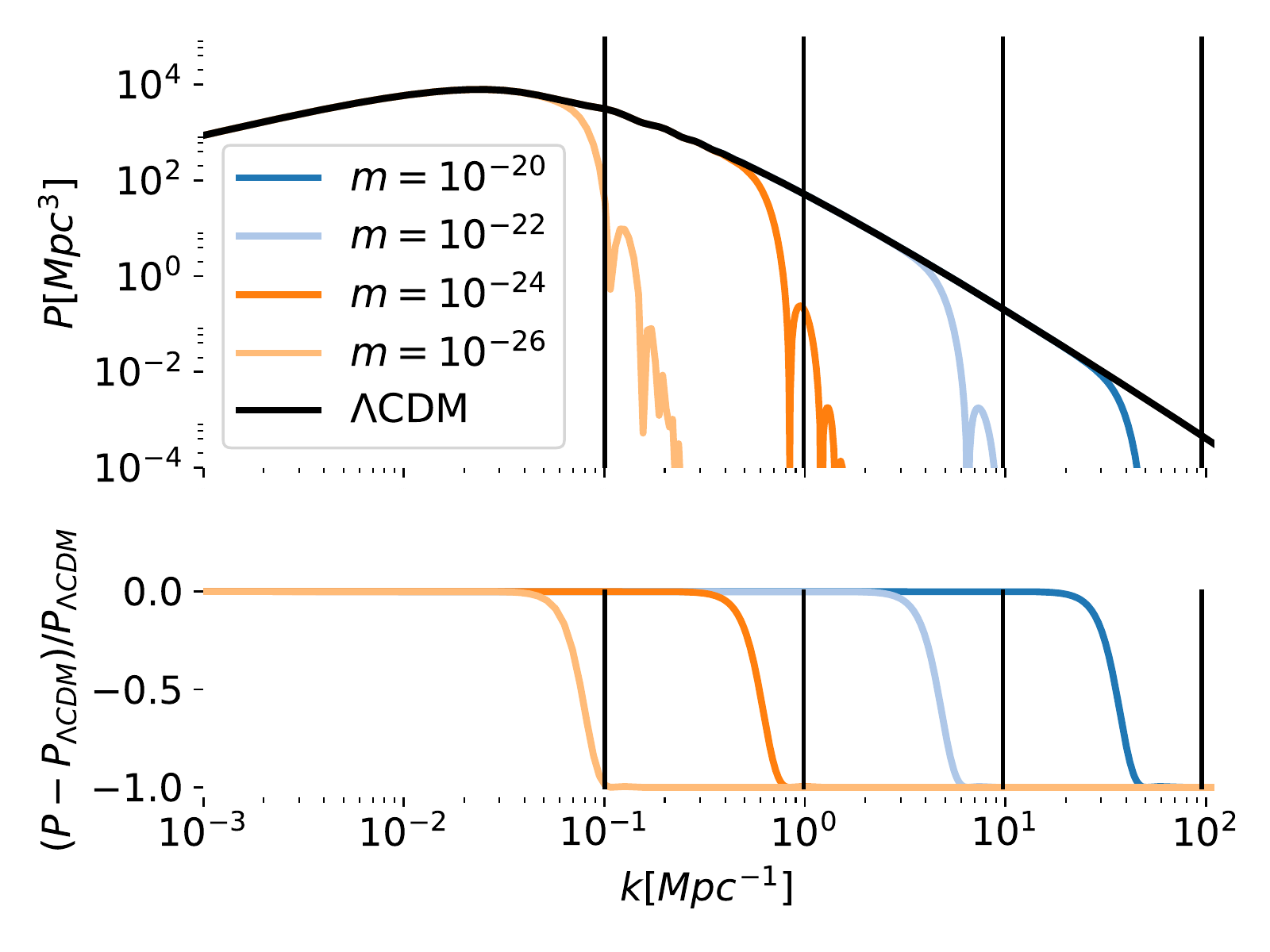}
\caption{Power spectrum for the  $\Lambda$SFDM  model for several masses of a Klein Gordon field. The coloured curves follow the prescription of Hu \textit{et~al.}~in \cite{Hu:2000ke} (the concordance $\Lambda$CDM power spectrum is shown in black). The cut-off predicted from Eq.~\eqref{k:th1} for a few SFDM models are represented by vertical lines reproducing the numerical values of Table I.}
\label{figure:1}
\end{figure}

While the exponent in the power law differs slightly ($4/9\sim 0.444$ in Hu's formula while $2/(5-n_s)\sim 0.496$ in our formula), the numerical values for the relevant dark matter models are quite similar. The specific values of $k_{\rm th}$ for a range of masses of SFDM are displayed in Table I. The comparison with the semianalytical values of the cutoff of the curvature power spectrum (given in \cite{Hu:2000ke}) are shown in Figure \ref{figure:1}. 
%for the masses considered in \cite{Urena-Lopez:2015gur}.
\begin{table}[h!]
  \label{table:cutoff}
\vskip-.5cm
  \begin{center}
  \begin{tabular}{cccc}
    \\
    Field Mass &$\displaystyle\frac{k_{\rm lin}}{\rm{Mpc}^{-1}}$& $\displaystyle\frac{\kth}{\rm{Mpc}^{-1}}$&$\displaystyle\frac{k_{\rm Hu}}{\rm{Mpc}^{-1}}$\\
    \hline
    \hline
    $10^{-20}$eV & 101&95.2 & 50.3 \\
    $10^{-22}$eV &10.1& 9.7 & 6.5 \\
    $10^{-24}$eV &1.01& 0.99 & 0.84 \\
    $10^{-25}$eV &0.32& 0.32 & 0.30 \\
    $10^{-26}$eV &0.10& 0.10 & 0.108  \\
    \\
  \end{tabular}
        \caption {Fourier modes for the cut-off in the power spectrum for several masses of the $\Lambda$SFDM model. The second column corresponds to the linear cut-off (\ref{cortelineal}), the
        third column corresponds to the non-linear threshold obtained in eq. (\ref{k:thl}),
        while the last column corresponds to the Hu \textit{et~al.}~ cut off obtained in ref. \cite{Hu:2000ke}
        and written in eq. (\ref{k:th3}).}
        %%\JDS{Habra que decir porque los valores no corresponden con los de Hu mas que a orden de magnitud?} 
  \end{center}
\end{table}

Our successful match of cut-off scales for the SFDM power spectrum provides a novel interpretation for the existence of this feature. The fluctuations that lie above the instability threshold $\zeta_{c}$ at horizon crossing given in Eq.~\eqref{crit:zeta:sf} will collapse, while those with a smaller amplitude will dissipate. While inaccuracies may be due to the fluctuations entering the horizon at the radiation era, the rough consistency with numerical studies gives us confidence to look further into implications of our result at other cosmological stages dominated by oscillating scalar fields. In particular, the next subsection is devoted to the implications of our threshold amplitude in the formation of structure at reheating stages. 

%%%%%%%%%%%%%%%%%%%%%%%%%%%%%%%%%%%%%%%%%%%%
\subsection{Primordial Black Holes in a reheating scenario}
\label{sec:ivB}
%%%%%%%%%%%%%%%%%%%%%%%%%%%%%%%%%%%%%%%%%%%%

In the cosmological stages following the inflationary era, the energy stored in the inflationary field(s) must be transfered to the standard model fields by the time of Big Bang Nucleosynthesis. Reheating models seek an effective process for such energy transfer. One of such models, preheating, relies on resonant oscillations between the inflaton and other matter fields, e.g. a second scalar field (for a recent review see \cite{Amin:2014eta}) to reach an efficient energy transfer. This is usually modeled with the inflaton oscillating at the bottom of its potential, for which a good approximation is the Klein-Gordon field $V(\varphi) = m^2\varphi^2/2$. In large field models, the mass of the field can be inferred from the normalization of the amplitude of fluctuations measured by the Planck satellite. The standard value is roughly $m \simeq 3.4\times 10^{12}\,{\rm GeV}$ (see e.g.~\cite{Jedamzik:2010dq}). The oscillatory phase can last for a significant period until the Universe reaches an equilibrium temperature $T_{\rm rh}$, at the thermalisation time, giving way to the standard Hot Big Bang. 
\iffalse 
In order to avoid the overproduction of unstable gravitinos and reach a stable BBN, the thermalisation temperature must be below $T_{\rm rh} \lesssim 3\times 10^{8}\,\mathrm{GeV}$. 
\fi
It is thus not unrealistic to consider a long period when the matter density is dominated by the oscillating inflaton, and where the condition $H/m \ll 1$ is met, so that our results in the previous sections are valid. In particular, from Eq.~\eqref{crit:zeta:sf} the threshold amplitude for curvature perturbations to collapse and form structures could be of order
\beq
\label{zetac:reheat}
\zeta_{c} = 10^{-16},
\eeq
%%% This is the value of zeta at T= 10^11\,GeV and m = 1.4*10^-6 form zeta3.nb 
%%% but we must review it in terms of zeta2-alt.nb.
 
\noindent for an oscillating inflaton at the energy scale\footnote{\setstretch{0.9} Even lower values than \eqref{zetac:reheat} for the threshold amplitude $\zeta_c$ can be obtained if the oscillating inflaton dominates the matter density at scales below $E \lesssim 10^{8}\,\mathrm{GeV}$.}
$E = 10^{10} \,\mathrm{GeV}$.
Comparing this value to the primordial amplitude of Eq.~\eqref{primordial:zeta}, one would infer that most of the primordial inhomogeneities should collapse gravitationally in this era. The instability of fluctuations at the preheating era result in the formation of bound structures, solitons or boson stars of the oscillating field called ``oscillons'' \cite{Amin:2014eta}. Numerical simulations are consistent with the formation of such objects which quickly dominate the energy budget. Oscillons subsequently decay into relativistic particles and their effects impact observables at several subsequent stages. Long-living oscillons can delay the thermalisation and extend the number of e-foldings in the reheating phase. Oscillons can also interact and generate an observable Gravitational Waves background.  The threshold proposed in this paper could be useful to count the matter density in terms of solitons and to compute numbers of solitons with the Press-Schechter formula. 

Looking at the formation of primordial black holes (PBHs) in the reheating environment, previous works have shown that the conservative approach of adopting the radiation era threshold for collapse ($\zeta_c \gtrsim 0.71$) could overproduce PBHs in some particular preheating models \cite{Torres-Lomas:2014bua}. However, contrary to the situation of a universe dominated by a fluid with a hard equation of state, in an environment with negligible average pressure not all of the inhomogeneities above the critical value would collapse. Black hole formation is prevented by the deviations from the spherical symmetry of fluctuations. The geometry of inhomogeneities can be computed from the random variables that parametrise the amplitude of matter density perturbations in each spatial direction. A black hole will only form when all of the mass $M$ of a given inhomogeneity lies within a region internal to a ball of radius smaller or equal to the associated Schwarzschild radius $4\pi GM/c^2$. In Ref. \cite{Harada:2016mhb}, a recent calculation of this requirement results in a formula for the initial matter density fraction of PBHs $\beta(M)$ 
 \beq
 \label{beta:sigma}
 \beta(M_{BH}) = b_0 \, \sigma^5(M) \,,	
 \eeq
\noindent where $\sigma(M)$ is the variance of the matter fluctuations of mass $M$ and the coefficient $b_0$ was
determined numerically as $b_0=0.056$ or semi-analytically in the range $0.1280 < b_0 < 0.01338$. PBHs are formed
with a fraction $\gamma$ of the Hubble mass at the time of horizon entry, for which the associated radius is
the comoving Hubble scale $r_H = (aH)^{-1}$. The variance is, on the other hand, related to the primordial power spectrum
$\mathcal{P}_\zeta$ by \cite{Josan:2009qn}
 \beq
 \label{sigma:powerspectrum}
 \sigma^2(M) = \frac{16}{3}\int (kr_H)^2  j_1^2\left(\frac{kr_H}{\sqrt{3}}\right)e^{-k^2r_H^2}\mathcal{P}_\zeta (k) \frac{dk}{k},
 \eeq
 \noindent where $j_1(x)$ is the spherical Bessel function and where we have assumed a Gaussian window function.
 As a result, the dominant contribution to the integral comes from wave numbers $k\sim 1/r_H$.
 For a primordial power spectrum of the form (\ref{primordial:zeta}) the integral can be performed
 analytically in terms of the generalized hypergeometric functions and the $\Gamma$ function as
 \begin{equation}
   \sigma^2 = \frac{2}{9}\frac{A_s}{(r_H/20 \rm{Mpc})^{n_s-1}}I \,.
 \end{equation}
 where
 \begin{equation}
    I =\Gamma\left( \frac{3+n_s}{2} \right)
    \prescript{}{2}{F}_2 \left(\left\{ \frac{3}{2},\frac{3+n_s}{2}\right\},\left\{ 2,3 \right\},-\frac{1}{3}   \right) \,.
 \end{equation}
 \noindent Substituting the central value of the spectral index reported by Planck \cite{Ade:2015xua} of $n_s=0.9681$ 
% this expression simplifies to
% \begin{equation}
%    \sigma^2 = 4.020402\times 10^{-10}\left(\frac{R}{20 \rm{Mpc}} \right)^{0.0319} \,.
% \end{equation}
the factor $I$ yields the numerical value $I = 0.84498$.

The above results are a function of the black hole mass $M_{BH}$ which is related
to the Hubble mass at the time of horizon crossing by $M_{BH}=\gamma M_H$. This in turn is related
to the {comoving} Hubble scale $r_H$ by $M_H=4\pi(ar_H)^3\rho/3$.
We consider that during reheating the oscillating scalar field
dominates the energy density of the Universe evolving, as we've seen, as a pressureless perfect fluid on average, $\rho=\rho_{\rm rh}(a/a_{\rm rh})^{-3}$, where the subindex ${}_{\rm{rh}}$ indicates the quantities evaluated at
thermalisation, corresponding to the end of the reheating period and the onset of the radiation dominated period. 
% \begin{equation}
%    R= \left(\frac{3M_H}{4\pi\rho} \right)^{1/3} \frac{1}{a} \,.
% \end{equation}
% The values of the density $\rho_{\rm rh}$ and scale factor $a_{\rm rh}$ can be evaluated at any epoch during the reheating period,
%not only the period of formation of the black holes.
%in particular we consider both quantities to be
%evaluated at the end of reheating and the onset of the radiation dominated period of the universe.

Assuming an instantaneous transfer of energy from the field to the radiation density, and the subsequent
conservation of total entropy in the radiation component, we can relate the thermalisation parameters to the current radiation values as
 \begin{equation}\label{reheat_density}
    a^3_{\rm rh} \rho_{\rm rh} = a_0^3 \rho_{r0} \frac{T_{\rm rh}}{T_{r0}} \,.
 \end{equation}
 %Finally, we consider $\gamma$ as the ratio between the mass of the Black Hole and the Hubble mass $M_{BH}=\gamma M_H$.
 Combining Eqs. (\ref{beta:sigma})-(\ref{reheat_density}) we can write
 an expression for the fraction of primordial black holes formed at the thermalisation time as
 \begin{equation}\label{beta_simp}
    \left(\frac{\beta}{ b_0}\right)^{2/5} =\frac{2}{9} A_s I
    \left[ \left( \frac{3M_{BH}}{4\pi\gamma} \frac{T_{r0}}{\rho_{r0}T_{\rm rh}} \right)^{1/3} k_0
             \right]^{1-n_s}.
 \end{equation}
% where $I=0.845$ for an $n_s=0.9681$.
 One of the main goals in the study of PBHs is to constrain the fraction
 of these objects as a function of its mass in order to impose bounds on the primordial power spectrum at small scales \cite{Carr:2009jm,Josan:2009qn}. However, the parameter to be constrained is usually dependent on the fraction of the Hubble mass to the black hole mass $\gamma$,
 and the number of degrees of freedom of the radiation during the formation of the black hole $g_{*i}$. Therefore, it is customary to constrain the rescaled mass fraction parameter $\beta'(M)$ defined as:
 \begin{equation}
    \beta'(M_{BH}) = \gamma^{1/2}\left( \frac{g_{*i}}{106.75} \right)^{-1/4} \beta(M_{BH}) \,.
 \end{equation}
 Assuming that the black holes are formed close to the thermalisation time, and adopting
 central values of $n_s$ from Planck, we obtain an expression for the mass fraction in terms of the black hole mass
 \bea
     \beta' &=& 3.16\times 10^{-27} \frac{b_0}{0.056} \times \notag \\
     &&\left( \frac{g_*}{230} \right)^{-0.24}
     \left( \frac{\gamma}{0.2} \right)^{0.46}
     \left( \frac{M_{BH}}{10^{15}\mathrm{gr.}} \right)^{0.040}
    \,.
    \label{eq:reheating_predictions}
 \eea
 This expression can be contrasted against the set of observational constraints at the relevant masses. It is also useful
 to write this expression in terms of the thermalization temperature using the relation
 \begin{equation}
    \frac{M_{BH}}{10^8 \rm GeV} =
    1.287
     \frac{\gamma}{0.2}  \sqrt{\frac{230}{g_*}}
     \left( \frac{T_{\rm rh}}{10^8\rm{GeV}} \right)^{-2}
     \,.
  \end{equation}
  
  \noindent Equation (\ref{eq:reheating_predictions}) thus becomes
 % \begin{equation}\label{eq:reheating_predictions}
 %   \beta' = 3.1\times 10^{-27} \left( \frac{M_{BH}}{10^{15}g} \, \frac{10^8 GeV}{T_{\rm rh}} \right)^{0.02658}
 %   \frac{b_0}{0.056}\,.
 %\end{equation}
 %This expression is better written in terms of the thermalization temperature only or the black hole mass only
 %we write both expressions here:
  \bea
     \beta' &=& 3.20\times 10^{-27} \frac{b_0}{0.056}\times \notag \\
     &&\left( \frac{g_*}{230} \right)^{-0.26}
     \left( \frac{\gamma}{0.2} \right)^{1/2}
     \left( \frac{T_{\rm rh}}{10^8\rm{GeV}} \right)^{0.080}
    \,. \label{eq:betaT}
 \eea
 
Let us now show that for black holes in the mass range $10^{12}$ to $10^{16}$ grams produced at thermalisation, the value of $\beta'$ inferred from the sphericity condition is very small, nevertheless it is still comparable to current constraints from
cosmological and astrophysical observations. In figure \ref{fig:betas} we compare the predictions of black hole
production from Eq. (\ref{eq:reheating_predictions}) to up-to-date constraints in the relevant mass range: From Ref. \cite{Tashiro:2008sf} the fraction of black holes of mass $10^{11}$ to $10^{13}\,\mathrm{gr.}$ is constrained to be
$\beta'>10^{-21}$ otherwise the evaporating black holes before recombination would have noticeable effects in the CMB thermal spectrum (blue horizontal line in Fig. \ref{fig:betas}). Meanwhile, in the mass range $2.5\times10^{13}\,\mathrm{gr.} < M_{BH} < 2.4\times 10^{14}\,\mathrm{gr.}$
the black holes will evaporate between the epoch of recombination and up to redshift $z \approx 6$, radiating a small
fraction of the mass in electrons and positrons which can damp the high $l$ CMB anisotropies, unless its fraction
is smaller than $\beta' < 3\times 10^{-30}({M_{BH}}/{10^{13}\,\mathrm{gr.}})^{3.1}$ as reported in \cite{Carr:2009jm}. For black holes
with mass greater than $2.5\times 10^{13}\,\mathrm{gr.}$ another constrain comes from the fact that their radiation can contribute to the
diffuse x-ray and $\gamma$-ray backgrounds, according to
Carr \textit{et al.}~\cite{Carr:2009jm} the black holes that would evaporate {between the recombination time and the current epoch},
which correspond to  $2.5\times 10^{13} \,\mathrm{gr.} <M_{BH} < M_*$ with $M_*=5.1\times 10^{14}\,\mathrm{gr.} $
should approximately satisfy $\beta'<3\times10^{-27}(M_{BH}/M_*)^{-2.9}$ while the black holes with masses above $M_*$
which are yet to evaporate completely should satisfy  $\beta'<4\times10^{-26}(M_{BH}/M_*)^{3.3}$.
Furthermore for black holes in the range $10^{15}$ to $10^{17}$ grams,  Ref. \cite{Clark:2016nst} computed an
update on the effects of the evaporating PBHs on the history of reionization and on the damping of
CMB anisotropies obtaining tighter bounds than those from the $\gamma$-ray background given by
$\beta'<2.4\times 10^{-26} (M_{BH}/M_*)^{4.3}$.

All the above constraints are plotted in figure \ref{fig:betas} together with the predictions from the black hole mass
fraction $\beta'(M_{BH})$ produced at a range of admissible thermalisation energies. We see that the observational constraints are violated for black holes produced in an interval of masses from $2.5\times10^{13}\,\mathrm{gr.} $ up to around $10^{14}\,\mathrm{gr.} $ and also for PBHs of mass just below $5.1\times10^{14} \,\mathrm{gr.}$

Our analysis requires confirmation that the sphericity condition is the most stringent criterion of PBH formation in a reheating environment. This should come from numerical simulations of PBH formation under such conditions. If the sphericity condition results in smaller values of $\beta(M_{BH})$, our results then rule out thermalisation at energy scales compatible with the featured masses, that is in the range $3.75\times 10^8 \leq T_{\rm rh}\leq 7.18\times 10^8$ and at the value $T_{\rm rh} \approx 1.59\times 10^8 \, \mathrm{GeV}$ \footnote{Another possibility is that an excess of tiny PBHs produced after the end of inflation may trigger the reheating process through the evaporation products \cite{Hidalgo:2011fj}}.
%%\textbf{What is that interval?}
 Our results constrain the reheating period reducing the admissible energy densities at which a model with an oscillating K-G field dominates the early universe. Obviously, future observations will improve the bounds to PBH abundance and consequently constrain further the reheating scenarios.
 \vskip-0.5cm
{\centering
\begin{figure}[t!]
%\begin{center}
\includegraphics[width=9cm]{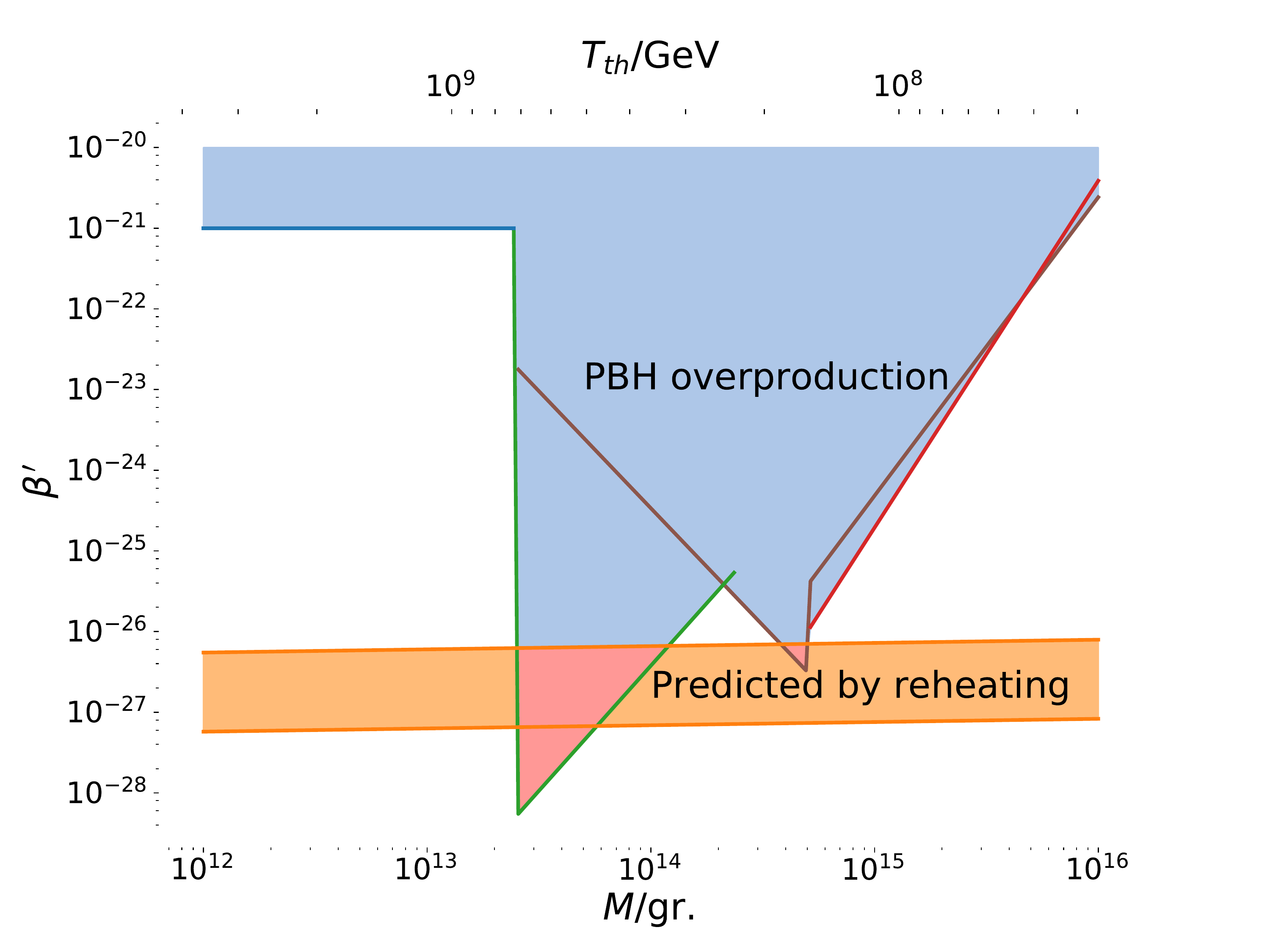}
\caption{Mass fraction of primordial black holes at the time of formation. The orange bar corresponds to the prediction from the sphericity condition, valid for an oscillating inflaton, while
the blue and red regions are forbidden by observational constraints. There are two intervals where
the reheating scenario saturates the observational bound. The constraints come from observations of the
CMB thermal spectrum (blue horizontal line), damping of large $l$ CMB anisotropies (green line), direct detection
of x-rays or $\gamma$-rays from evaporating black holes (brown lines), and modification on the reionization history
and CMB power spectrum (red line) as described in the text.}
\label{fig:betas}
%\end{center}
\end{figure}
}
%%%%%%%%%%%%%%%%%%%%%%%%%%%%%%%%
%VALORES: Rango figura $4 \times 10^7$ a $4 \times 10^9$. Rango de Temperaturas de intersección $3.75\times 10^8$ --$7.18\times 10^8$. and 
%$1.59\times 10^8 \mathrm{GeV}$.
%%%%%%%%%%%%%%%%%%%%%%%%%%%%%%%%
  \section{Summary and Discussion}
\label{sec:V}

In this paper we have used the well-known instability scale of a cosmological Klein-Gordon scalar field to derive the threshold amplitude for the collapse of its overdensities. We arrived at this threshold following the spherical collapse model, and in particular borrowing the methods to derive the threshold for Primordial Black Hole (PBH) formation in barotropic fluid environments \cite{Carr:1975qj,Harada:2013epa}. 
We have checked the validity of our result by correctly reproducing the well-known cut-off of the matter power spectrum in the Scalar Field Dark Matter model (SFDM), also dubbed fuzzy or Ultra-Light Axion Dark Matter. Specific values of our analytic cutoff are contrasted with those derived from the perturbative treatment in Figure \ref{figure:1} showing reasonable accuracy. 

This first result is promising. Our analysis can be extended to the study of structure formation in more intricate dark matter models with a clear instability scale. For example, more convolved Axion Dark Matter models which predict gravitational instabilities analogous to the Jeans wavelength. Our analysis provides a straightforward method to derive the cut-off of the matter powerspectum, and furthermore, an estimate of the mass fraction of scalar field dark matter in the form of collapsed objects in the context of the Press-Schechter formalism. 

%Our analytic expression in Eq.~(24) can be used further to explore the impact of such 
%cut-off in complementary observables such as number counts in the context of the 
%Press-Schechter formalism.

The consistency of the threshold amplitude motivates us to consider the structure formation in the reheating scenario. Recent articles looking at PBH formation, inspired by the detection of gravitational waves by pairs of black holes, prompt us to look at the probability of primordial black hole formation employing our threshold amplitude for collapse. We have found, however, that the structures resulting from gravitational instability are oscillons and the collapse to a black hole is prevented by the geometry of the collapse, in a situation equivalent to that of a matter component with negligible average pressure. Our threshold value is therefore not directly applicable to compute the probability of Primordial Black Hole formation (The threshold of PBH formation is yet to be determined for a universe dominated by an oscillating scalar field). The formation of PBHs is, however, subject to the sphericity condition, which yields very low values for the mass fraction of PBHs $\beta(M)$.
\iffalse
\footnote{\setstretch{0.9} At higher energies, the PBH formation can be dictated by this threshold instead of the sphericity condition. \textit{Tenemos que hacer una figura al respecto}}. 
\fi

We have computed the probability of PBH formation $\beta(M)$ for PBHs formed in a K-G field (reheating) scenario, in the black hole mass range $10^{12}$ to $10^{16} \mathrm{gr.}$  This corresponds to thermalisation scales of an oscillating K-G field in the range $ 8 \times 10^{9} \gtrsim T_{\rm rh}/ \mathrm{GeV} \gtrsim 3\times 10^{7}$. For some of the masses in this range, PBHs are produced more abundantly than the associated observational constraint (mostly violating observations of CMB spectral distortions). 

It is important to note the lack of numerical studies of PBH formation from scalar fields, and the consequent absence of a PBH formation threshold to determine the PBH mass fraction precisely. Developing a relativistic simulation of PBH formation during reheating is the subject of future work. At the moment, following the sphericity condition seems to be the only alternative to compute the formation of PBHs in a universe dominated by oscillating scalar fields. If the resulting $\beta(M)$ stands as the most stringent value for PBH mass fraction, the overproduction of PBHs displayed in Figure \ref{fig:betas} would rule out the existence of an oscillating scalar field thermalising the universe at energy temperatures of order ${ T_{\rm rh} = (3.7-7.2) \times{10^{8}}\,\mathrm{GeV}}$. 

\iffalse
\begin{itemize}
  \item Mention that a higher (lower) reheating temperature would
    increase  (reduce) the value of $\beta_{\rm PBH}$ since the more
    time PBHs spend in a radiation era, the larger the relative
    density becomes.  
    \item Mention that even if the reheating temperature is considered too low (down to the BBN scales of 100 MeV), the bounds from PBHs to oscillating scalar fields at different energy scales are not 
\end{itemize}
\fi

{\em Acknowledgements:} 
The authors are grateful to Karim Malik and Alexander Polnarev for useful discussions.  We gratefully acknowledge support from \textit{Programa de Apoyo a Proyectos de Investigaci\'on e Innovaci\'on Tecnol\'ogica} (PAPIIT) UNAM,  IA-103616. We are also thankful to CONACYT for support from research grants CONACYT/239639 and CONACYT/269652.

%%%%%%%%%%%%%%%%%%%%%%%%%%%%%%%%%%%%%%%%%%%%%%%%%%%%%%%%%%%%%%%%%%%%
%%%%%%%%%%%%%%%%%%%%%%%%%%%%%%%%%%%%%%%%%%%%%%%%%%%%%%%%%%%%%%%%%%%%

\end{document}